\documentclass[11pt, reqno]{amsart}
\usepackage{geometry}                
\usepackage{graphicx}
\usepackage{amssymb}
\usepackage{epstopdf}

\newtheorem{theorem}{Theorem}

\title{The Initial Value Problem in General Relativity\footnote{To appear as a Chapter of "The Springer Handbook of Spacetime," edited by A. Ashtekar and V. Petkov. (Springer-Verlag, at Press)}}
\author{James Isenberg}

\date{}                                            

\begin{document}
\maketitle

\section{Introduction}
\label{Intro}

Ever since Newton's formulation of particle mechanics over three hundred years ago, one of the most widely used methods of modeling physical systems is via an initial value formulation. For particle mechanics, the idea is that one specifies the initial position $\vec y_0$ and initial momentum $\vec p_0$ of the particle, and one then determines the particle's path $\vec y(t)$ in time by solving the initial value problem consisting of Newton's equations (an ordinary differential equation system) 
\begin{eqnarray} 
\nonumber
\frac{d }{dt}\vec y(t) &=&\frac{1}{m} \vec p(t), \\
\nonumber
\frac{d}{dt}\vec p(t) &=&\vec F(\vec y, \vec p),
\end{eqnarray}
together with the initial conditions $\vec y(t_0)=y_0$ and $\vec p(t_0)=p_0$. Analogously, to study the vibrational motion  $\psi (x,t)$ of a stretched string, one specifies the initial displacement $\psi_0(x)$ and initial velocity $\nu_0(x)$ of the string, and one then determines the subsequent motion $\psi(x,t)$ by solving  the initial value problem consisting of the (first order) wave equation (a partial differential equation system) 
\begin{eqnarray}
\nonumber
\partial_t \psi(x,t) &= &\nu(x,t), \\
\nonumber
\partial_t \nu(x,t) & =& \alpha \partial_{xx} \nu(x,t),
\end{eqnarray}
together with the initial conditions $\psi(x,t_0)= \psi_0(x)$ and $\nu(x,t_0)= \nu_0(x)$.

Although one of the signature properties of Einstein's theory of gravity is its spacetime covariant character, it too admits an initial value formulation. We recall that to model the gravitational interactions of physical systems, general relativity incorporates the physics of the gravitational field into the geometry of spacetime, which is specified by a four-dimensional manifold $M^4$ together with a metric $g$ of signature $(-,+,+,+)$. Einstein's equations relate the spacetime metric to the non-gravitational ``matter fields" $\Psi$ which are present in the spacetime by requiring that  the Einstein tensor  $G_{\mu\nu} := R_{\mu\nu} - (1/2) R g_{\mu\nu}$ of $g$ (here $R_{\mu\nu}$ is the Ricci tensor of $g$ and $R$ is its scalar curvature)  be proportional to the stress-energy tensor $T_{\mu\nu}$ of $\Psi$ and $g$; specifically, 
\begin{equation}
\label{Einsteq}
G_{\mu \nu}[g] = \kappa T_{\mu \nu}[g, \Psi],
\end{equation} 
where $\kappa= 8\pi \mathcal{G}_{N}$ with $\mathcal{G}_{N}$ Newton's gravitational constant.
One of the most  effective ways to construct and study a solution $(M^4, g, \Psi )$ to Einstein's equations \eqref{Einsteq} is by specifying a set of initial data corresponding to an initial ``state of the universe", and then using Einstein's equations (reformulated as an initial value problem) to evolve this data into the corresponding  spacetime solution. This chapter discusses the details of this initial value formulation for Einstein's theory: the specific nature of the initial data sets for which the initial value problem works, how one finds such initial data sets, how one reformulates Einstein's equations \eqref{Einsteq} as an initial value problem and verifies that it is well-posed, and some open problems related to the evolution of initial data sets.

One very important feature of the initial value formulation of Einstein's theory is that the initial data must satisfy a set of \emph{constraint equations}. This is a familiar feature of Maxwell's theory of electromagnetism as well: An initial data set for Maxwell's theory consists of a pair of spatial vector fields $(\vec E_0(\vec x),\vec B_0(\vec x))$
which are required to satisfy the Maxwell constraint equations 
\begin{eqnarray}
\nonumber
\nabla \cdot \vec E_0 = 0,\\
\nonumber
\nabla \cdot \vec B_0 = 0
\end{eqnarray}
(presuming that the charge density is zero).
A solution $(\vec E (\vec x,t),\vec B(\vec x,t))$ is obtained from the initial data  via the evolution equations\footnote{Here and throughout this chapter, we set $c$, the speed of light equal to one.}
\begin{eqnarray}
\nonumber
\partial_t \vec B (\vec x,t) &= &\nabla \times \vec E(\vec x,t),\\
\nonumber
\partial_t \vec E (\vec x,t) &= &-\nabla \times \vec B(\vec x,t). 
\end{eqnarray}
It is important to note that presuming a data set $(\vec E_0(\vec x),\vec B_0(\vec x))$ satisfies the Maxwell constraints, it follows from the evolution equations that the solution generated from this data set satisfies the constraints $\nabla \cdot \vec E(\vec x,t) = 0$ and $\nabla \cdot \vec B(\vec x,t) = 0$ for all times $t$.

To understand the initial value formulation of Einstein's theory and how it produces spacetimes  $(M^4, g, \Psi )$ which satisfy Einstein's equations, it is useful to start by examining spatial foliations of such a spacetime, with a focus on the geometric fields induced on the spatial leaves of such a foliation and how they relate to the spacetime metric $g$. We do this in Section \ref{3+1}. Included in this section is a discussion of the Gauss-Codazzi-Mainardi equations which relate the spacetime curvature of $g$ to the induced fields. Based on these equations,  we derive the constraint and evolution equations which comprise the initial value formulation of Einstein's theory, and then in Section \ref{Well-Posed} we discuss the proof that this  initial value problem is well-posed. Next, we examine methods for obtaining initial data sets which satisfy  the Einstein constraint equations. In Section \ref{Conformal} we focus on  the conformal method, in Section \ref{CThin} we discuss the closely related conformal thin sandwich method, and in Section \ref{Glue}, we explore  how such solutions can be obtained via gluing. Our discussion of the Einstein evolution equations is very brief; in Section \ref {Evoln} we comment primarily on globally hyperbolic solutions with singularities and the strong cosmic censorship conjecture, along with brief remarks concerning the stability of Minkowski spacetime, the Kerr solutions, and certain expanding cosmological solutions of the Einstein equations.

\section{$(3+1)$-foliations and the derivation of the Einstein constraint and evolution equations}
\label{3+1}

To see how to build spacetime solutions of the Einstein field equations from initial data on a Riemannian manifold, it is useful to start with such a spacetime $(M^4, g, \Psi )$, specify a spatial foliation of that spacetime (presuming that it admits such\footnote{So long as the spacetime is globally hyperbolic, as defined below, it admits a spatial foliation. Spacetimes
which are not globally hyperbolic do not generally admit spatial foliations.}),
determine the geometric quantities which are well-defined on the leaves of that foliation, and then calculate what the Einstein equations \eqref{Einsteq} tell us about these geometric quantities. We do this here. 

A three-dimensional hypersurface $\Sigma^3$, smoothly\footnote{To simplify the discussion and avoid issues of regularity, we presume that all manifolds and fields are smooth, by which we mean that they are infinitely differentiable.} embedded in $(M^4,g)$ (with embedding map $i:\Sigma^3 \rightarrow M^4$), is defined to be \emph{spacelike} (or \emph{spatial}) if the induced bilinear form $\gamma:=i^{*}g$ on $\Sigma^3$ is Riemannian. Equivalently, the embedded hypersurface is spacelike if all vectors $V$ tangent to $i(\Sigma^3)$ at $p$ are spacelike with respect to $g$ (so that $g(V,V)>0$). A \emph{spatial foliation} of the spacetime $(M^4,g)$ is a smooth one-parameter family $i_t :\Sigma^3 \rightarrow M^4$ (for  $t\in \mathbb{R}$)  of embedded spacelike hypersurfaces such that each point $p\in M^4$ is contained in one and only one of the hypersurfaces  $i_t(\Sigma^3)$ (referred to as the ``leaves" of the foliation). We presume that it is also true that for each $p\in M^4$, there is one and only one point $q\in \Sigma^3$ such that $i_{\hat t} (q)=p$ for some $\hat t$.

We focus now on a particular leaf $i_t (\Sigma^3)$ of a chosen foliation $i_t$ of a chosen spacetime $(M^4,g)$. The definition of spatial foliation (above) guarantees that $i_t (\Sigma^3)$ is equipped with a Riemannian metric $\gamma_{[t]}$, which consequently defines a covariant derivative $\nabla_{[t]}$ and the corresponding Riemann, Ricci, and scalar curvature quantities, all intrinsic to $i_t (\Sigma^3)$. In addition to specifying these intrinsic geometric quantities, the foliation $i_t$ equips  $i_t (\Sigma^3)$ with extrinsic geometric quantities, including (i) $e_{\perp [t]}$, a future-pointing unit-length timelike vector field, orthogonal to all vectors tangent to $i_t(\Sigma^3)$;  (ii) $\theta^{\perp}_{[t]}$, a unit-length one-form field such that (at each point in $i_t(\Sigma^3)$), $\langle \theta^{\perp}_{[t]}, e_{\perp[t]}\rangle=1$ and $\langle \theta^{\perp}(p), V\rangle=0$ for all vectors $V$ tangent to $i_t(\Sigma^3)$; and $K_{[t]}$, the second fundamental form, defined by
\begin{equation} 
\label{Kdef}
K_{[t]} (U, V) := g(D_U e_{\perp[t]}, V),
\end{equation}
where $D$ denotes the spacetime covariant derivative and where $U$ and $V$ are vector fields tangent to $i_t(\Sigma^3)$. It is straightforward to show that $K_{[t]}$ is a well-defined (spatial) tensor field with respect to  the tangent spaces of $i_t(\Sigma^3)$, and further that it is symmetric. By contrast, $e_{\perp[t]}$ and $\theta^{\perp}_{[t]}$ are not spatially  tensorial; rather, they act tensorially on the spacetime tangent spaces of $M^4$, restricted to the subset $i_t(\Sigma^3) \subset M^4.$

To relate the foliation-related quantities $\gamma_{[t]}, \nabla_{[t]}$, the spatial curvatures, $K_{[t]}, e_{\perp [t]}$ and $\theta^{\perp}_{[t]}$ to the spacetime geometry, it is useful to specify foliation-compatible bases. We obtain a (local in space) foliation-compatible coordinate basis by choosing coordinates $\{x_a\}| _{(a=1,2,3)}$ locally on $\Sigma^3$, using the foliation to transport these to its leaves $i_t(\Sigma^3)$, and adding the parameter $t$ to produce the spacetime coordinates $\{x_a,t\}=:\{x_{\alpha}\}|_ {(\alpha =1,2,3,0)}.$ The corresponding dual coordinate bases $\{\partial_{\alpha}\}$ and $\{dx^{\alpha}\}$ are foliation-compatible in the sense that $dt$ annihilates vectors which are tangent to the leaves, $\partial_t$ is transverse to the leaves, and $\partial_a$ are tangent to the leaves; however, we note that  $\partial_t$ and $dt$ need not be timelike\footnote{$\partial_t$ is timelike if $g(\partial_t, \partial_t)<0$ and it is null if $g(\partial_t, \partial_t)<0$; similarly $dt$ is timelike or null or spacelike depending on the sign of $g^{-1}(dt,dt)$.}
everywhere. Alternatively, one may choose the foliation-compatible basis $\{e_{\perp [t]}, \partial_a\}$ and its dual $\{\theta^{\perp}_{[t]}, \theta^a_{[t]}:=dx^a+M^adt\}$; here $M^a_{[t]}$ are the components of the \emph{shift vector}, which is the spatial projection of $\partial_t$:
\begin{equation}
\label{lapse&shift} 
\partial_t =N_{[t]}e_{\perp [t]} + M^a_{[t]}\partial_a. 
\end{equation}
The scalar $N_{[t]}$ is called the \emph{lapse function}.

On the leaf $i_t (\Sigma^3)$, one easily verifies that the spacetime metric $g$ is related to the spatial metric $\gamma_{[t]}$ by the identity (and its expansion)  
\begin{eqnarray}
\label{g&gamma}
g &=& \gamma_{[t]} - \theta^{\perp}_{[t]} \theta^{\perp}_{[t]}\\
    &=& \gamma_{[t] ab}(dx^a + M^a_{[t]}dt)(dx^b + M^b_{[t]}dt)-N^2_{[t]}dt^2.
\end{eqnarray}
The expression relating the spacetime covariant derivative $D$ (compatible with $g$) and the spatial covariant derivative $\nabla_{[t]}$ (compatible with $\gamma_{[t]})$ is not so easily derived; however it does follow from the definition \eqref{Kdef} of the second fundamental form that for any pair of vectors $U$ and $V$ tangent to $i_t(\Sigma^3)$, one has 
\begin{equation}
\label{D&nabla}
D_U V = \nabla_{[t] U} V + K_{[t]}(U,V) e_{\perp [t]},
\end{equation}
thus identifying $K_{[t]}$ with the surface-normal projection of the spacetime covariant derivative. One also verifies from this definition, as well as from the Lie derivative identity $\mathcal{L}_Y g_{\alpha \beta} =D_{\alpha} Y_{\beta} + D_{\beta}Y_{\alpha}$, that 
\begin{equation}
\label{metrictimederiv}
\mathcal{L}_{\partial_t} \gamma_{[t] ab} =2N_{[t]} K _{[t] ab} + \mathcal{L}_{M_{[t]}} \gamma_{[t]ab}.
\end{equation}
Notably, this result involves not only quantities on the slice $i_{[t]}(\Sigma^3)$, but also the rate of change of one of these quantities--$\gamma_{[t]}$--along the foliation.

It remains to consider the relationship between the spacetime curvature and the spatial curvatures. Since the calculations
which lead to formulas for the components of the spacetime curvature tensor in terms of components of the spatial curvature tensor along with other quantities intrinsic to leaves of a chosen foliation are analogous to analyses first done (for Riemannian rather than for Lorentzian ambient geometries) by Gauss, Codazzi and later Mainardi, these formulas are often referred to as the Gauss-Codazzi-Mainardi equations. Using 
\begin{equation}
\label{SptmCurv}
\mathcal{R}^{\alpha}_{\ \beta \gamma \delta}:=\langle \theta^{\alpha}, D_{e_{\gamma}} D_{e_{\delta}} e_{\beta} - D_{e_{\delta}} D_{e_{\gamma}} e_{\beta} - D_{[e_{\gamma}, e_{\delta}]} e_{\beta} \rangle
\end{equation} 
to represent the spacetime curvature tensor (with respect to spacetime dual bases $\{e_{\alpha}\}$ and $\{\theta^{\beta}\}$) and analogously 
\begin{equation}
\label{SpCurv}
R^a_{\ bcd}:= \langle \theta^a, \nabla_{e_c} \nabla_{e_d} e_b - \nabla_{e_d} \nabla_{e_c} e_b- \nabla_{[e_c, e_d]} e_b \rangle
\end{equation}
to represent the spatial curvature tensor (with respect to spatial dual bases $\{e_a\}$ and $\{\theta^b\}$), we find that these equations take the following form\footnote{The most straightforward method of deriving the Gauss-Codazzi-Mainardi equations is by combining  the expression \eqref{D&nabla} relating $D$, $\nabla$ and $K$ with the expressions \eqref{SptmCurv} and \eqref{SpCurv} for the curvatures. Details of these calculations appear in \cite{IN80}.}:
\begin{equation}
\label{Gauss}
\mathcal{R}^a_{\ bcd} = R^a_{\ bcd}+K^a_c K_{bd} - K^a_d K_{bc},
\end{equation}
\begin{equation}
\label{Codazzi}
\mathcal{R}^a_{\ \perp cd} = \nabla_{e_c}K^a_d - \nabla_{e_d}K^a_c,
\end{equation}
and 
\begin{equation}
\label{Mainardi}
\mathcal{R}^a_{\ \perp b \perp} = -\mathcal{L}_{e_{\perp}}K^a_b +K^{am}K_{mb} + \frac{1}{N} \nabla^a \nabla_b  N.
\end{equation}
Note that, to avoid notational clutter, in equations \eqref{SpCurv}-\eqref{Mainardi} we have removed  the ``$[t]$" subscripts from foliation-defined quantities such as $\nabla_{[t]}$ and $K_{[t]ab}$.

If the spacetime $(M^4, g, \Psi )$ under consideration satisfies the Einstein field equations \eqref{Einsteq} and if a spatial foliation $i_t:\Sigma^3 \rightarrow M^4$ has been specified, what do these equations tell us about the foliation-defined geometric quantities $\gamma_{[t]}$ and $K_{[t]}$? Combining the definition of the Einstein tensor $G_{\alpha \beta}=\mathcal{R}_{\alpha \beta} - \frac{1}{2}\mathcal{R} g_{\alpha \beta}$ with the Gauss-Codazzi-Mainardi equations \eqref{SpCurv}-\eqref{Mainardi}, we readily obtain the following
\begin{eqnarray}
\label{HamConstr}
R+K_{cd}K^{cd} - (K_c^c)^2 = \frac{1}{2} G_{\perp \perp} = \frac{\kappa}{2} T_{\perp \perp} [\gamma, \psi, \pi], \\
\label{MomConstr}
\nabla_c K^c_d -  \nabla_d (K^c_c)  = G_{\perp d} = \kappa T_{\perp d} [\gamma, \psi, \pi], \\
\label{Evoln}
\mathcal {L}_{e_{\perp}} K_{ab} + R_{ab} -2 K_{ac}K^c_b + K^c_c K_{ab} +\frac{1}{N} \nabla_a \nabla_b N &=&G_{ab}-\frac{1}{2} g_{ab} G^c_c \\
\nonumber &=& \kappa T_{ab}[\gamma, \psi, \pi] - \frac{\kappa}{2}g_{ab} T^c_c [\gamma, \psi, \pi], 
\end{eqnarray}
where $(\psi, \pi )$ represent the non-gravitational fields and their derivatives, after they have been decomposed into quantities which are well-defined with respect to the leaves of the foliation.\footnote{For Maxwell's electromagnetic field, $(\psi, \pi)$ correspond to $(B_a, E_a).$}

The first two sets of these equations \eqref{HamConstr}-\eqref{MomConstr} have the very interesting feature that (presuming the quantity $T_{\perp \perp} [\gamma, \psi, \pi]$ behaves well), they involve quantities ($\gamma_{[t]}$ and $K_{[t]}$, not $e_{\perp}$) defined as tensors on the leaf $i_t(\Sigma^3)$, and do \emph{not} directly involve any time derivatives of these quantities. By contrast, the remaining set of these equations \eqref{Evoln} does involve the time derivative of $K_{[t]}$. 

There is nothing special about $i_t(\Sigma^3)$ or any other leaf of the foliation. Nor is there anything special about the (arbitrarily) chosen foliation. Hence we see that on any spacelike hypersurface embedded in a spacetime $(M^4,g, \Psi)$ which satisfies the Einstein equations, the surface quantities $(\gamma, K, \psi, \pi)$ must satisfy \eqref{HamConstr}-\eqref{MomConstr}. These are hence known as the \emph{initial value constraint equations} for Einstein's theory.

The initial value problem (or Cauchy problem) for Einstein's theory turns things around. It addresses the following question: If we fix a three-dimensional manifold $\Sigma^3$ and if we choose a Riemannian metric $\gamma_{ab}$ and a symmetric tensor $K_{cd}$ (and possibly, non-gravitational field data $\psi$ and $\pi$ on $\Sigma^3$  as well) which satisfy the constraint equations \eqref{HamConstr}-\eqref{MomConstr}, is it always true that there exists a spacetime $(M^4, g, \Psi)$ which satisfies the (spacetime) Einstein field equations \eqref{Einsteq}, and which contains an embedded submanifold $i(\Sigma^3) \subset M^4$ for which $\gamma_{ab}$ is the induced metric $i^{*}g_{ab}$ and $K_{cd}$ is the induced second fundamental form? Further, if such a solution exists for a given set of initial data, is it unique in any sense? We address this question in the next section.

\section{Well-posedness of the initial value problem for Einstein's equations}
\label{Well-Posed}

It appears, from our discussion in Section \ref{3+1}, that the construction of a (vacuum) solution of the Einstein equations from initial data is a straightforward enterprise (especially with a computer readily at hand): a) On a specified three-dimensional manifold $\Sigma^3$, we choose an initial data set $(\gamma_{ab}, K_{cd})$ which satisfies the (vacuum) constraint equations
\begin{eqnarray}
\label{HamConstrVac}
R+K_{cd}K^{cd} - (K_c^c)^2 = 0 \\
\label{MomConstrVac}
\nabla_c K^c_d -  \nabla_d (K^c_c)  = 0;
\end{eqnarray}
b) we freely choose the lapse as a one-parameter ($t$) family  of positive scalar functions $N(x,t)$ on $\Sigma^3$, and the shift as a one-parameter family of vector fields $M^a(x,t)$ on $\Sigma^3$;   c) using the \emph{evolution equations}
\begin{eqnarray}
\label{EvolnMetric}
\mathcal{L}_{\partial_t} \gamma_{ab}& = &2N K _{[t] ab} + \mathcal{L}_{M} \gamma_{ab}, \\
\label{EvolnK}
\mathcal {L}_{\partial_t} K_{ab}& =& - N R_{ab} + 2N K_{ac}K^c_b  - N K^c_c K_{ab} - \nabla_a \nabla_b N +\mathcal{L}_M K_{ab},
\end{eqnarray}
we evolve the initial data set into a one-parameter family of Riemannian metrics $\gamma_{ab}(x,t)$ and symmetric tensors $K_{ab}(x,t)$; and 
d) we construct the Lorentz metric 
\begin{equation}
g= \gamma_{ab}(x,t) (dx^a + M^a (x,t)dt)(dx^b + M^b(x,t) dt)-N^2(x,t)dt^2
\end{equation}
on the spacetime manifold $M^4= \Sigma^3 \times \mathbb{R}$ and verify that it is a solution of the vacuum Einstein equations $G_{\alpha \beta}=0.$

To determine that this construction procedure in fact works, at least for some choices of the lapse and shift, one needs to prove a well-posedness theorem. Such a theorem was first proven for Einstein's theory during the 1950's by Yvonne Choquet-Bruhat \cite{FB52}. This early work leaves the issue of uniquenness unsettled to a certain extent. However, the later work of Choquet-Bruhat with Robert Geroch \cite{CBG69} provides  a strong form of uniqueness. Before stating this combined result, we find it useful to establish some terminology:

A spacetime $(M^4,g)$ is \emph{globally hyperbolic} if there exists an embedded spatial hypersurface $i(\Sigma^3) \subset M^4$ such that every future and past inextendible causal path\footnote{A smooth path $\eta:I\rightarrow M^4$ is \emph{causal} if its tangent vector $V$ is always either timelike ($g(V,V)<0$) or null (g(V,V) = 0). Such a path is \emph{inextendible} if there does not exist a smooth path $\tilde{\eta}$ which contains $\eta$ as a proper subset.} intersects $i(\Sigma^3)$ once and only once. If such a hypersurface exists, it is called a \emph{Cauchy surface}. A spacetime is a \emph{globally hyperbolic development} (``gh-development") of a specified set of initial data $(\gamma, K)$ on $\Sigma^3$ if (i) it as a solution of the Einstein field equations; (ii) it is globally hyperbolic; and (iii) there exists an embedded hypersurface 
$\tilde i(\Sigma^3)$ such that $(M^4,g)$ induces the specified initial data $(\gamma, K)$ on $\tilde i(\Sigma^3)$ in the sense that $\gamma= \tilde i ^{*} g$ and an appropriately modified version of \eqref{Kdef} holds. A spacetime $(M^4,g)$ is a \emph{maximal globally hyperbolic development} of $(\gamma,K)$ on $\Sigma^3$ if every other gh-development of the same set of data is diffeomorphic to a subset of $(M^4,g)$. 

Using this terminology, we have the key result \cite{FB52,CBG69}:

\begin{theorem} [Well-posedness of Einstein's vacuum equations] 
\label{WPThm} 
For any smooth set of initial data $(\gamma, K)$ on $\Sigma^3$ which satisfies the vacuum constraint equations \eqref{HamConstrVac}-\eqref{MomConstrVac}, there exists a unique (up to diffeomorphism) maximal globally hyperbolic development.
\end{theorem}

Before discussing how to prove this result and discussing ways in which the result can be generalized, we note a number of its features: First, Theorem \ref{WPThm} is fundamentally a local existence and uniqueness result. It guarantees that for any initial data set, there exist spacetimes which are gh-developments of that data, and there exists a unique maximal gh-development, but it says nothing regarding whether that maximal development lasts for a finite or an infinite amount of proper (or coordinate) time. Second, Theorem \ref{WPThm} guarantees that there are globally hyperbolic spacetimes evolving from any given set of initial data, but it says nothing about non-globally hyperbolic spacetimes containing an embedded hypersurface 
with that data. In many cases---the Taub-NUT spacetime \cite{HE73} and the various generalized Taub-NUT spacetimes \cite{M84} are prominent examples---the maximal development of a given set of data can be extended smoothly across a Cauchy horizon
 to a non-globally hyperbolic spacetime solution. A Cauchy horizon in a given spacetime $(M^4,g)$ (\emph{not} to be confused with a Cauchy surface, as discussed above) is a hypersurface $\mathcal{H}$ embedded in $(M^4,g)$ which is null (i.e., at each point of $\mathcal{H}$, there is a vector tangent to $\mathcal{H}$ which is null) and which lies on the boundary between the region in $(M^4,g)$ which is globally hyperbolic and the region which is not. The initial value problem tells us essentially nothing about whether or not such extensions exist for a given set of initial data.
Third, Theorem \ref{WPThm} guarantees that there are choices of the lapse and shift which can be used in evolving initial data sets to produce gh-developments (numerical or otherwise), but it does not tell us whether this is true just for certain particular choices, or for all such choices. The proof of Theorem \ref{WPThm}, as we see below, does provide some information regarding this issue, but does not resolve it. Finally, we note that while well-posedness theorems usually contain statements regarding continuity of the map from the domain space of initial data sets to the range space of solutions, this is not true of Theorem \ref{WPThm}, as stated. Such a result does hold in an appropriate form for Einstein's equations; however, to avoid the unenlightening detail needed to state it, we do not include it here. We refer the interested reader to Chapter 15 of \cite{R09}. 

The key for proving that the initial value problem for Einstein's field equations is well-posed is to show that the system \eqref{Einsteq} can be, in a certain sense, transformed into a hyperbolic PDE system\footnote{ A PDE system is \emph{hyperbolic} if, roughly speaking, it has the characteristics of a system of wave equations. A precise definition of  hyperbolic PDE systems is found, for example,  in \cite{E10}, and also in \cite{R09}.}
for which well-posedness is well-established. The vacuum Einstein equation system, which we can write as $\mathcal{R}_{\alpha \beta} =0$, is not itself 
hyperbolic. The system 
\begin{equation}
\label{ModEin}
\mathcal{R}_{\alpha \beta} + \frac{1}{2} \mathcal{L}_Z g_{\alpha \beta} =0,
\end{equation}
with 
\begin{equation}
Z_{\alpha} := -g^{\nu \mu}(\partial_{\nu}g_{\mu \alpha} - \frac{1}{2} \partial_{\alpha} g_{\nu \mu}) + \mathcal{F}_{\alpha}[g]
\end{equation}
for $\mathcal{F}_{\alpha}[g]$ a specified function of $g_{\alpha \beta}$ but \emph{not} of its derivatives, takes the form 
\begin{equation} 
g^{\nu \mu} \partial_{\nu}\partial_{\mu} g_{\alpha \beta} = \mathcal{Q}_{\alpha \beta} [g, \partial g],
\end{equation} 
and therefore \emph{is} a hyperbolic PDE system for the metric $g_{\alpha \beta}$. This does not lead to the Einstein system itself being hyperbolic. However, using the Bianchi identities, one can show that the vector field $Z^{\alpha}$ satisfies the homogeneous hyperbolic system
\begin{equation} 
\label{Zeqn}
g^{\nu \mu} \partial_{\nu}\partial_{\mu} Z^{\alpha} = - \mathcal {R}^{\alpha}_{\beta}Z^{\beta},
\end{equation}
so that the combined system of \eqref{ModEin}-\eqref{Zeqn} is hyperbolic. Further, one can show (see, for example, Chapter 14 of \cite{R09}, or Section 3.1 of \cite{BI04}) that, so long as the geometric initial data $(\gamma, K)$ satisfy the constraints, one can choose the corresponding initial values of $g_{\alpha \beta}$ and of  $\partial_t g_{\alpha \beta} $ for the system \eqref{ModEin}-\eqref{Zeqn} in such a way that the initial values of $Z^{\alpha}$ and $\partial_t Z^{\alpha}$ vanish. It then follows from standard results regarding (nonlinear) hyperbolic systems (see, for example,  Chapter 9 of \cite{R09}) that the system  \eqref{ModEin}-\eqref{Zeqn} is well-posed; further, it follows that the solution $(g_{\alpha \beta}(x,t), Z^{\alpha}(x,t))$ has $Z^{\alpha}(x,t)$ vanishing for all time. Hence $(g_{\alpha \beta}(x,t), 0)$ is a solution of the vacuum Einstein equations, compatible with the initial data $(\gamma, K)$.

We note that in the original proof \cite{FB52} of the well-posedness of the Einstein vacuum system, wave coordinates (then called ``harmonic coordinates") are used. Wave coordinates correspond to a particular choice of the function  $\mathcal{F}_{\alpha}[g]$. Generalized wave coordinates, which play an important role in numerical simulations of black hole collisions \cite{P05} correspond to other choices of $\mathcal{F}_{\alpha}[g]$. 

The techniques used to show that, for any given set of initial data $(\gamma, K)$ satisfying the constraint equations, there exists a unique maximal spacetime development of that data, are very different from those (just discussed) which are used to prove local existence. To prove that a unique maximal development exists, the idea is to consider the set $\mathcal{M}_{[\gamma, K]}$ of all spacetime developments of $(\gamma,K)$, and then define a partial ordering ``$>$" on $\mathcal{M}_{[\gamma, K]}$, with $(\tilde M^4,\tilde g) > (M^4,g)$ if $(\tilde M^4, \tilde g)$ is an extension of $(M^4,g)$, up to diffeomorphism. One then shows that the ordered set $(\mathcal{M}_{[\gamma, K]},>)$ has the properties needed to apply the Maximality Principle from set theory\footnote{The Maximality Principle is equivalent to the Axiom of Choice. It is often referred to as ``Zorn's Lemma, although some dispute the appropriateness of this label (see Chapter 16 of \cite{R09}).}, from which it follows that $(\mathcal{M}_{[\gamma, K]},>)$ has a unique maximal element. One readily verifies that this maximal element is indeed a maximal spacetime development of the data $(\gamma,K)$. Details of this argument can be found in \cite{CBG69} and in Chapter 16 of \cite{R09}.

There are two important ways in which one can generalize Theorem \ref{WPThm}. The first of these involves coupling in non-gravitational fields. It is \emph{not} true that if a field theory has a well-posed initial value problem in Minkowski spacetime, then the standard (``comma-to-semicolon"\footnote{Roughly speaking, the``comma-to-semicolon" coupling of a given field theory to gravity involves replacing partial derivatives with metric-compatible covariant derivatives in the Lagrangian for the given field theory, multiplying this Lagrangian by the metric volume density $\sqrt{-det g}$, adding the result to the Einstein Lagrangian $\mathcal{R}\sqrt{-det g}$, and then varying the summed Lagrangian with respect to the metric and the fields.}) 
coupling of that field theory to Einstein's theory  also necessarily has a well-posed initial value problem. The standard Klein-Gordon vector field theory, with field equations $D^{\alpha}D_{\alpha} W^{\mu} =m^2 W^{\mu}$, provides an example of a field theory which is well-posed in flat spacetime, yet appears to be ill-posed if coupled to Einstein's theory.\footnote{It is generally difficult to \emph{prove} that a system of partial differential equations does not have a well-posed Cauchy problem. However, for systems like the Klein-Gordon vector field theory with standard coupling to Einstein's theory, the  presence of ``derivative-coupling" terms in the equations cause standard analyses of hyperbolicity and well-posedness to be essentially unmanageable; one is led to strongly suspect ill-posedness. Some of these issues are discussed in \cite{IN77}.}
However, there are a number of field theories involving Einstein's theory coupled to non-gravitational fields that do have well-posed Cauchy problems. These include Einstein-Maxwell, Einstein-Yang-Mills, Einstein-Dirac\footnote{The verification of well-posedness is straightforward for the Einstein-Dirac theory with commuting spinor fields. For Einstein-Dirac with anti-commuting spinor fields, and also for $N=1$ supergravity (which requires that the spinor fields anti-commute), well-posedness holds, but in a subtle sense. See \cite{BCBIY85} for details.}, certain forms of Einstein-scalar field theories, and certain forms of Einstein-fluid theories. For all of these, in addition to well-posedness, one can prove that for each set of initial data satisfying the appropriate set of constraints, there is a unique maximal development spacetime solution of the coupled system.

The second way in which Theorem \ref{WPThm} can be generalized is by loosening the required degree of regularity. The original results, as well as the results we state here, presume that the initial data is smooth. Results of over forty years ago \cite{FM72, HKM77} show that well-posedness (as well as the existence of a unique maximal development) holds for initial data sets $(\gamma, K)$ with $\gamma$ contained in a local Sobolev space\footnote{The Sobolev index $s$ indicates (weak) $L^2$ boundedness for order $s$ derivatives of the indicated fields. See, e.g., \cite{E10} for  definitions and properties of the Sobolev function spaces.} of index $s>5/2$, and $K$ in such a space with index $s-1$. The work of Klainerman and Rodnianski \cite{KR05}  lowers the required regularity to $s>2$, and the recent work of Klainerman, Rodnianski and Szeftel \cite{KRS12} indicates that (with certain restrictions) one can prove well-posedness for $s=2$ as well. We note that the drive to achieve lower regularity well-posedness results is motivated  both by the desire to understand weak solutions, and by the use of low regularity results as tools in understanding long-time behavior of solutions.

Not stated as part of Theorem \ref{WPThm} above, but an important feature of Einstein's theory, is the fact that the quantities $(\gamma_{ab}, K_{cd})$ induced on any Cauchy surface in a spacetime solution  of Einstein's equations must satisfy the constraint equations. Hence, as one evolves a spacetime solution from a set of initial data, the evolution equations \eqref{EvolnMetric}-\eqref{EvolnK} effectively preserve the constraints. 
Since numerical implementation of the Cauchy problem inevitably introduces small errors, the constraints are not precisely preserved in a numerically constructed solution. This appears to be a source of instability problems in numerical relativity.

\section{The conformal method of constructing and analyzing solutions of the constraint equations}
\label{Conformal}

The crucial first step in building a spacetime solution of the Einstein equations via the initial value problem is to obtain a set of initial data which satisfies the constraints. Restricting our attention to the vacuum system for most of the discussion here, we presume that a fixed three-dimensional manifold $\Sigma^3$ has been chosen, and we seek a Riemannian metric $\gamma_{ab}$ on $\Sigma^3$ and a symmetric tensor $K_{cd}$ on $\Sigma^3$ such that, together, $(\gamma_{ab},K_{cd})$ satisfy the vacuum constraints \eqref{HamConstrVac}-\eqref{MomConstrVac}. 

There are three somewhat different goals one might have in studying solutions of the constraint equations: i) the construction of physically interesting data sets, which may then be evolved into physically interesting spacetimes; ii) the parametrization (in terms of an appropriate function space) of the set of all solutions of the constraint equations; and iii) the systematic study of various mathematical and physical properties of solutions of the constraints, including (for asymptotically Euclidean data sets) conserved quantities such as global mass and global angular momentum.

The most widely used analytical method for studying solutions of the constraints is the conformal method (together with the closely related conformal thin sandwich method). The conformal method is particularly adapted to the goal of finding a function space parametrization of the set of solutions of the constraints, the second goal listed above. It has also been used very effectively in the numerical construction of physically interesting initial data sets, especially those used to simulate black hole collisions (which are important for building gravitational signal templates). We discuss the conformal method here and the conformal thin sandwich method in Section \ref{CThin}; in Section \ref{Glue}, we discuss another approach  for studying the constraints---gluing---which is especially effective as a tool for attaining the third goal listed above.

As a PDE system for the twelve functions encompassed in $\gamma_{ab}(x)$ and $K_{ab}(x)$, the four constraint equations \eqref{HamConstrVac}-\eqref{MomConstrVac} constitute an underdetermined system. The idea of the conformal system is to split the initial data into two sets---the ``free (conformal) data", and the ``determined data"--in such a way that, for a specified choice of the free data, the constraint equations become a \emph{determined} elliptical PDE system, to be solved for the determined data. There a number of ways to carry out this data split. We focus here on the ``semi-decoupling split" (labeled in the early literature  of the conformal method as ``method A"; see \cite{CBY80}), which takes the following form:

\begin{description}
\item[Free (``Conformal'') Data]\ %
\par

\begin{tabular}{rcl}
  $\lambda_{ab}$ &--& a Riemannian metric,
specified up to conformal factor;
\\
$\sigma_{ab}$ &--& a divergence-free\footnotemark \
($\nabla ^a\sigma_{ab} = 0$),
trace-free ($\lambda^{ab}\sigma_{ab} =0$)
\\
&& symmetric tensor;
\\
 $\tau$ &--& a scalar field;

\end{tabular}
\footnotetext{In the conformal data,
the divergence-free condition is defined using the Levi-Civita covariant
derivative compatible with the conformal metric $\lambda_{ab}$.}%
\item[Determined Data]\ %
\par

\begin{tabular}{rcl}
$\phi$ &--& a positive definite scalar field;
\\
   $W^a$ &--& a vector field.
\end{tabular}
\end{description}
\vspace{.1 in}
For a given choice of the free data, the four equations to be solved for the four functions of the determined data take the form 
\begin{eqnarray}
\nabla_m (LW)^m_a &=& \frac{2}{3}\phi^6\nabla_a \tau 
\label{confmom}
\\
\Delta\phi &=& \frac{1}{ 8}R\phi - \frac{1}{ 8}(\sigma^{mn} +
LW^{mn})(\sigma_{mn} + LW_{mn})\phi^{-7}  + \frac{1}{ 12}\tau^2 \phi^5,
\label{Lichnerocoupled}
\end{eqnarray}
where the  Laplacian $\Delta$ and the scalar curvature $R$ are
based on the $\lambda_{ab}$-compatible covariant derivative
$\nabla_a$, and where $L$ is the corresponding conformal Killing
operator, defined by
\begin{equation}
\label{LW}
(LW)_{ab}:=\nabla_aW_b +\nabla_bW_a -\frac{2}{3} \lambda_{ab} 
\nabla_m W^m.
\end{equation}
 Presuming that for the chosen conformal  data one can indeed solve equations (\ref{confmom})-(\ref{Lichnerocoupled}) for $ \phi$ and $W$, then the initial data set $( \gamma_{ab}, K_{cd})$ constructed via the formulas
\begin{eqnarray}
\gamma_{ab} &=& \phi^4 \lambda_{ab}, 
\label{recomposemetric}\\
K_{ab} &=& \phi^{-2}(\sigma_{ab} + LW_{ab} ) +\frac{1}{3} \phi^4 \lambda_{ab} \tau, 
\label{recomposeK}
\end{eqnarray}
satisfies the Einstein constraint equations \eqref{HamConstrVac}-\eqref{MomConstrVac}.

The four equations \eqref{confmom}-\eqref{Lichnerocoupled}, which we collectively refer to as the LCBY equations (since their derivation is based on the work of Lichnerowicz \cite{L44}, Choquet-Bruhat, and York \cite{CBY80}), are readily obtained by substituting the formulas \eqref{recomposemetric}-\eqref{recomposeK} into the vacuum constraints \eqref{HamConstrVac}-\eqref{MomConstrVac}. Two key identities play a major role in the
 derivation of \eqref{confmom}-\eqref{Lichnerocoupled}: The first is the formula for the scalar curvature of the conformally-transformed metric $\gamma_{ab}= \phi^4 \lambda_{ab}$, expressed in terms of the scalar curvature for $\lambda_{ab}$ and derivatives of $\phi$:
\begin{equation}
R(\gamma) = \phi^{-4} R(\lambda) -8 \Delta_{\lambda} \phi.
\end{equation}
We note that if we were to use a different power of $\phi$ as the conformal factor  multiplying $\lambda_{ab}$, then this formula would involve squares of first derivatives of $\phi$ as well. The second key formula relates the divergences of a traceless symmetric tensor $\rho_{ab}$ as calculated using two different covariant derivatives---$\nabla_{(\gamma)}$ (compatible with the metric $\gamma$) and $\nabla _{(\lambda)}$ (compatible with $\lambda$):
\begin{equation}
\nabla_{(\gamma)}^m \rho_{mb} = \phi^{-2} \nabla_{(\lambda)}^m (\phi^2 \rho_{mb}).
\end{equation}
This formula dictates the choice of conformal scaling used in \eqref{recomposeK} for the trace-free part of $K_{ab}$; a different scaling would generally lead to the appearance of further $\nabla \phi$ terms in the LCBY equations, which are to be avoided.

Do the LCBY equations admit solutions $(\phi, W)$ for every choice of the conformal  data $(\Sigma^3; \lambda_{ab}, \sigma_{cd}, \tau)$? 
It is easy to see that this is not the case: If we choose, for example, $\Sigma^3= S^3$, $\lambda_{ab}$ is the round metric on $S^3$, $\sigma_{cd}=0$ everywhere, and $\tau=1$ everywhere, then \eqref{confmom} takes the form $\nabla_m (LW)^m_a=0$, which requires that $LW_{ab}$ vanish everywhere.  Equation \eqref{Lichnerocoupled} then takes the form $\Delta \phi = \frac{1}{8} R \phi +\frac{1}{12} \phi^5$.  Since the right hand side of this equation is positive definite (recall the requirement that $\phi>0$), it follows from the maximum principle on closed (compact without boundary) manifolds that there is no solution.

Since this example shows that solutions do not exist for every possible choice of the conformal data, we seek to determine exactly which sets of such data lead to a solution and which do not. This issue has been intensively studied for at least forty years, and while much is known, there remains much to be determined. Roughly speaking, for conformal data with constant mean curvature (CMC), 
we generally know for which sets of conformal data $(\Sigma^3; \lambda_{ab}, \sigma_{cd}, \tau)$ solutions exist, and for which sets they do not. Uniqueness is well understood as well for CMC conformal data. For
conformal data with nearly constant mean curvature, we know a number of classes of conformal data for which solutions exist, as well a number of classes for which they do not; there are, however, a number of unresolved cases. For conformal data with mean curvature far from constant, we know much less; however, there has been some progress recently in studying these sets of conformal data. 

To explain more specifically what is known and what is not known about solving the LCBY equations, it is very useful  to classify conformal data sets into a wide collection of classes, based on a number of criteria, which include the following: 
\begin{itemize}
\item \emph{Manifold type and asymptotic conditions}: Data on closed manifolds, asymptotically Euclidean data, asymptotically hyperbolic data, asymptotically conical data, asymptotically cylindrical data, or data on manifolds with boundary.
\item \emph{Regularity conditions}: Analytic data, smooth data, or data contained in specified Sobolev or H\"older spaces.
\item \emph{Coupled non-gravitational fields}: Vacuum Einstein, Einstein-Maxwell, Einstein-Dirac, Einstein-scalar, Einstein-fluid, or Einstein-Vlasov.
\end{itemize}

It is  beyond the scope of this article to discuss what is known for each of the several classes delineated by these criteria. Rather, we  focus on what is known and what is not known for smooth data satisfying the vacuum Einstein constraints, either on closed manifolds or satisfying  asymptotically Euclidean conditions. We include  comments on some of the other classes at the end of this section. 

Within each class of conformal data, the key distinction is between those sets of data with constant mean curvature and those which have non-constant mean curvature. The mean curvature of any initial data set is given by the function $tr K:= \gamma^{ab}K_{ab}=\tau$ (we use \eqref{recomposemetric}-\eqref{recomposeK} to calculate the last equality here), so we see that the CMC condition corresponds to choosing conformal data with constant $\tau$.  Examining \eqref{confmom}, we see that the CMC condition is important because constant $\tau$ implies the vanishing of the right hand side of equation \eqref{confmom}, which then results\footnote{It follows from a straightforward analysis of the elliptic operator $\nabla \cdot L$ that in essentially all cases, if $\nabla_m(LW)^m_a=0$, then $LW_{ab}$ vanishes.}  in the vanishing of the tensor quantity $(LW)_{ab}$. Consequently, for CMC conformal data the analysis of the LCBY equations reduces to that of the (decoupled)  \emph{Lichnerowicz equation}, which takes the form 
\begin{equation}
\label{Lichnero}
\Delta\phi = \frac{1}{ 8}R\phi - \frac{1}{ 8}\sigma^{mn} \sigma_{mn} \phi^{-7}  + \frac{1}{ 12}\tau^2 \phi^5.
\end{equation}
We note that for all of the non-gravitational fields coupled to gravity which are listed above, the first of the LCBY equations takes the form $\nabla_m (LW)^m_a = \frac{2}{3}\phi^6\nabla_{\partial_a} \tau +J_a$, where $J_a$ depends on the non-gravitational conformal data\footnote{This effect occurs as a consequence of the choice one makes for the conformal rescaling of the non-gravitational fields. Alternative choices could be made which would lead to $J_a$ depending on $\phi$, but there is little motivation for making such choices.}, and does $\emph{not}$ involve $\phi$. Consequently, while $LW_{ab}$ does not vanish for CMC data if non-gravitational fields are being considered, the LCBY equations \emph{do} decouple, and as a result the determination if solutions to the LCBY equations exist depends essentially on the Lichnerowicz equation (with extra terms corresponding to the non-gravitational fields\footnote{For example, in the case of the Einstein-Maxwell theory, the Lichnerowicz equation picks up a term of the form $- \rho \phi^{-3}$, where $\rho$ is quadratic in the conformal electric and magnetic fields.}).

\subsection{CMC Data on Closed Manifolds}
\label {CMC-Closed}
As noted above, for a set of constant mean curvature conformal data $(\Sigma^3; \lambda_{ab}, \sigma_{cd},\tau)$ for the vacuum Einstein equations, a solution to the LCBY equations exists (and a map to a solution of the constraint equations exists) if and only if there exists a solution to the Lichnerowicz equation \eqref{Lichnero}. For this problem, we know exactly which sets of conformal data lead to solutions and which do not. To state these results (and to prove them), two features regarding conformal transformations of the conformal data are crucial.

First, we note the Yamabe theorem \cite{S84} for Riemannian metrics on closed manifolds, which states that every such metric can be conformally transformed to a metric of constant scalar curvature; further, for each metric the sign of that constant scalar curvature is unique. Hence, the set of Riemannian metrics on a given manifold $\Sigma^3$ are partitioned into three ``Yamabe classes" $\mathcal{Y}^+, \mathcal{Y}^0$, and $\mathcal{Y}^-$, depending on that sign.

Second, we readily verify that the Lichnerowicz equation is \emph{conformally covariant} in the following sense: If $\phi$ is a solution to the Lichnerowicz equation for a set of conformal data $(\Sigma^3; \lambda_{ab}, \sigma_{cd},\tau)$, then for any smooth positive function $\theta$, the function $\hat \phi = \theta ^{-1}\phi$ is a solution to the Lichnerowicz equation for the conformal data $(\Sigma^3; \theta^4 \lambda_{ab}, \theta^{-2}\sigma_{cd},\tau)$. Combining this result with the Yamabe theorem, we see that \emph{a solution to the Lichnerowicz equation exists for $(\Sigma^3; \lambda_{ab}, \sigma_{cd},\tau)$ if and only if there exists a solution for the conformally transformed data $(\Sigma^3; \theta^4 \lambda_{ab}, \theta^{-2}\sigma_{cd},\tau)$ with $R[\theta^4\lambda_{ab}]= +1,0$ or $-1$, depending upon the Yamabe class of $\lambda_{ab}$}. 

Besides the Yamabe class of $\lambda_{ab}$, two aspects of a given set of conformal data determine whether or not a solution to the Lichnerowicz equation exists. One of them is whether or not the constant $\tau$ is zero or not. In the table of results below, we label these alternatives $\tau=0$ or $\tau\not=0.$ The other is whether
or not the function $\sigma_{cd}\sigma^{cd}$ is identically zero (on $\Sigma^3$) or not. We label these alternatives $\sigma \equiv0$ or $\sigma \not \equiv 0$. Using ``Y" to indicate that solutions exist for conformal data sets in a certain class and using ``N" to indicate that they do not, we summarize the results in the following table:

\bigskip
\begin{tabular}{|c||c|c|c|c|}
\hline
&$\sigma \equiv 0,\tau=0$&$\sigma \not \equiv 0,\tau=0$&$\sigma \equiv 0,\tau\not =0$&$\sigma \not \equiv 0,\tau\not =0$\\
\hline
\hline
${\mathcal Y}^+$&N&Y&N&Y\\
\hline
${\mathcal Y}^0$&Y&N&N&Y\\
\hline
${\mathcal Y}^-$&N&N&Y&Y\\
\hline
\multicolumn{5}{c}{}\\
\end{tabular}
\bigskip

\noindent We note that for all those data sets such that solutions exist, except for that class of data with $\lambda \in \mathcal{Y}^0$, with $\sigma \equiv0$ and with $\tau=0$, the solutions are unique. In the latter (somewhat trivial) case, any (positive) constant $\phi$ is a solution.

The results summarized in the table here were to a large extent  proven in the 1970's by Yvonne Choquet-Bruhat, James York, and Niall O'Murchadha; see \cite{CBY80} for a discussion of this early work. The complete proof, including two cases not handled by the previous work, appears in \cite{I95}. There it is shown that the ``No" cases can all be proven using the \emph{Maximum Principle}, stated in the following form: On a closed manifold, the equation $\Delta \phi = f(x,\phi)$, with $f(x,\phi)$ either non-vanishing and non-positive or non-vanishing and non-negative, has no solution. It is also shown there that the ``Yes"  cases can all be proven using the \emph{Sub and Super Solution Theorem}, which can be stated as follows: If there exist a pair of positive functions $\phi_+ \geq \phi_-$ which satisfy the inequalities
\begin{equation}
\Delta\phi_- \leq \frac{1}{ 8}R\phi_- - \frac{1}{ 8}\sigma^{mn} \sigma_{mn} \phi_-^{-7}  + \frac{1}{ 12}\tau^2 \phi_-^5
\end{equation}
and 
\begin{equation}
\Delta\phi_+  \geq \frac{1}{ 8}R\phi_+ - \frac{1}{ 8}\sigma^{mn} \sigma_{mn} \phi_+^{-7}  + \frac{1}{ 12}\tau^2 \phi_+^5,
\end{equation}
then there exists a solution $\phi$ of the Lichnerowicz equation \eqref{Lichnero}, with $\phi_+ \geq \phi \geq \phi_-$. For some of the six ``Yes" cases (specifically those with the metric in the negative Yamabe class), constant sub and super solutions are easily found. For others, non-constant sub and super solutions are needed, and are not so easily found. 

We can use the results summarized in the table, along  with the conformal covariance stated above and a certain scaling invariance, to \emph{parametrize} the set of CMC solutions of the vacuum constraints on a chosen closed manifold $\Sigma^3$.  We note first that every solution $(\gamma_{ab}, K_{ab})$ of the constraint equations \eqref{HamConstrVac}-\eqref{MomConstrVac} can be obtained using the conformal method, since one can choose $\lambda_{ab}=\gamma_{ab}$, $\tau=K^c_c$, and $\sigma_{cd}$ equal to the divergence-free trace-free projection\footnote{As discussed in \cite{CBY80}, such a projection always exists} of $K_{cd}$; with this conformal data, $W^a=0$ and $\phi=1$ satisfy the LCBY equations, leading us back to $(\gamma_{ab}, K_{ab})$. Next, we note that the conformal covariance result implies that for any positive function $\theta$ the two sets of conformal data  $(\Sigma^3; \lambda_{ab}, \sigma_{cd},\tau)$ and $(\Sigma^3; \theta^4 \lambda_{ab}, \theta^{-2}\sigma_{cd},\tau)$ lead to related solutions of the Lichnerowicz equation and thence to \emph{identical} solutions $(\gamma_{ab}, K_{cd})$ of the constraint equations \eqref{HamConstrVac}-\eqref{MomConstrVac}. The scaling result \cite{BO10} states that for a positive constant $A$, the two sets of CMC conformal data $(\Sigma^3; \lambda_{ab}, \sigma_{cd},\tau)$ and $(\Sigma^3;  A^2 \lambda_{ab}, A \sigma_{cd}, A^{-1}\tau)$ admit the same solution $\phi$ of the Lichnerowicz equation. These lead to different solutions $(\gamma_{ab}, K_{cd})$ and $(A^2 \gamma_{ab}, A K_{cd})$ of the constraint equations; but these solutions are related by an essentially trivial scaling. As a consequence of these considerations, we determine that for any chosen closed manifold $\Sigma^3$, the set of all conformal data sets marked ``Yes" in the table, quotiented out by the conformal covariance equivalence and by the scaling equivalence, provides a faithful (bijective) parametrization of the space of CMC solutions of the vacuum constraints on $\Sigma^3$.

These results for CMC solutions of the vacuum constraints are essentially replicated if one considers the Einstein-Maxwell, Einstein-Yang-Mills, Einstein-Dirac, or Einstein-fluid\footnote{For the Einstein-fluid constraints, there are consistent choices of conformal scaling of the fluid fields which lead to a more difficult set of LCBY equations, analogous to those arising for certain Einstein-scalar theories. Conformal scalings can, however, always be chosen in a way which avoids these problems \cite{IMaxP05}.} constraint equations \cite{IMaxP05}. On the other hand, the results for the Einstein-scalar field theories with certain types of field theory potentials are not nearly as complete. Details of the difficulties that can arise are discussed in \cite{CBIP07}.

\subsection{Asymptotically Euclidean CMC Data}
\label{CMC-AE}
The decoupling of the LCBY equations which results from the assumption of constant mean curvature does not
depend upon the topology of $\Sigma^3$ or the asymptotic conditions of the conformal data. Hence, as for the closed manifold case discussed above, the analysis of the conformal method for asymptotically Euclidean data (as well as for asymptotically hyperbolic data or any other chosen asymptotic condition) reduces to the study of the solvability of the Lichnerowicz equation. We note that the fall-off conditions which define a set of data to be asymptotically Euclidean (see, for example, \cite{CBIY00} for a precise statement of these fall-off conditions) require that the mean curvature, if constant, be zero. Hence asymptotically Euclidean data which is CMC must be maximal; i.e., $K^c_c =\tau=0.$

The criterion for the Lichnerowicz equation to admit a solution for asymptotically Euclidean conformal data is independent of $\sigma_{cd}$, and  depends only the metric. As proven by Cantor \cite {C77}, the Lichnerowicz equation admits a solution for a given set of asymptotically Euclidean conformal data if and only if the metric admits a conformal transformation which results in the scalar curvature vanishing everywhere. Such metrics are called ``Yamabe positive", and Brill and Cantor \cite{BC81} have also shown (with a correction pointed out by Maxwell \cite{M04}) that an asymptotically Euclidean metric is Yamabe positive in the sense just described if and only if for every non-vanishing compactly supported function $f$ on $\Sigma^3$, the inequality 
\begin{equation}
\inf_{ \{ f \not\equiv 0\} }  \frac{\int_M (|\nabla f|^2 + R f^2) \sqrt{\det 
\lambda}}{||f||^2_{L^2}}>0.
\label{YamabeAE}
\end{equation}
holds.

We note that the same criterion for the solvability of the Lichnerowicz equation holds for asymptotically Euclidean data for the Einstein-Maxwell, Einstein-Yang-Mills, Einstein-Dirac, Einstein-fluid, and other such theories involving  non-gravitational fields coupled to Einstein's theory..

\subsection{Near-CMC Data}
\label{Near-CMC}
If we drop the non-CMC condition on the choice of the conformal data, then $\nabla \tau$ does not vanish, and we must deal with the fully coupled LCBY system.  While not much is known about the solvability of the LCBY system for general sets of non-CMC conformal data, for those sets of data with $|\nabla \tau|$ small in some appropriate sense, one can in many cases determine whether or not solutions exist. 

One way to analyze the coupled system is to use the iterated Gummel method, which replaces \eqref{confmom}-\eqref{Lichnerocoupled} by the sequence of semi-decoupled PDE systems
\begin{eqnarray}
\nabla_c (LW_{(n)})^c_a &=& \tfrac{2}{3}\phi_{(n-1)}^6\nabla_a\tau
\label{confmomA-n}
\\
\Delta\phi_{(n)}& =& \tfrac{1}{ 8}R\phi_{(n)} - \tfrac{1}{ 
8}(\sigma^{cd} +
LW_{(n)}^{cd})(\sigma_{cd} + LW_{(n)cd})\phi_{(n)}^{-7}  + \tfrac{1}{
   12}\tau^2 \phi_{(n)}^5. 
\label{LichneroA-n}
\end{eqnarray}
The idea is to (i) choose an initializing value for $\phi_0$; (ii) show that a sequence $(\phi_{(n)}, W_{(n)})$ of solutions to \eqref{confmomA-n}-\eqref{LichneroA-n} exists; (iii) prove that there are uniform upper and lower bounds for the sequence $(\phi_{(n)}, W_{(n)})$; and (iv) use those upper and lower bounds together with a contraction mapping argument to show that the sequence $(\phi_{(n)}, W_{(n)})$ converges uniformly to a limit $(\phi, W)$, which solves the LCBY equations. 

The assumption that $|\nabla \tau|$ is small plays a crucial role in carrying out steps (iii) and (iv). It does this because, as a consequence of the elliptic equation \eqref{confmom} and its sequential analog \eqref{confmomA-n}, the norm of $LW$  is controlled by $|\nabla \tau| \phi^6.$ Thence the term involving the square of $LW$ in \eqref{Lichnerocoupled}, which contains factors of $(\phi^6)^2$ times $\phi^{-7}$---i.e., $\phi^5$---competes with the $\tau ^2 \phi^5$ term. To ensure that $\tau^2 \phi^5$ (which has the favorable sign) wins this competition and to thereby retain the control of $\phi$ that one has in the CMC case, it is sufficient to require that $\frac{|\nabla \tau|}{|\tau|}$ be sufficiently small

Analyses of the sort discussed above have been carried out in \cite{IM96, ACI08}. Combining these results with others  proven in \cite{IO04} and \cite{DGH12}, one has the following table which lists, for each of twelve classes of near-CMC conformal  data on closed manifolds, whether solutions to the LCBY equations are known to exist\footnote{Uniqueness holds as well in all of these ``Y" cases.} (``Y"), are known to not exist (``N"), or are not known either way (``?"). In this table, the Yamabe classes and their labels $\{\mathcal{Y}^+, \mathcal{Y}^0, \mathcal{Y}^-\}$ are the same as used above  for  the CMC table, and the labels $\sigma \equiv 0$ and $\sigma \not \equiv 0$ denoting whether or not $\ |\sigma|$ is identically zero or not are also the same as for the CMC table. For the function $\tau$, the notation $\tau^2 > 0$ indicates those sets of conformal data with 
$\tau$ (presumed smooth) having no zeros, while $\tau^2 \not > 0$ labels those sets of conformal data with $\tau$ allowed to have zeroes.

\bigskip
\begin{tabular}{|c||c|c|c|c|}
\hline
&$\sigma \equiv 0,\tau^2 \not > 0$&$\sigma \not \equiv 0,\tau^2 \not >0$&$\sigma \equiv 0,\tau^2>0$&$\sigma \not \equiv 0,\tau^2 > 0$\\
\hline
\hline
${\mathcal Y}^+$&?&Y&N&Y\\
\hline
${\mathcal Y}^0$&?&Y&N&Y\\
\hline
${\mathcal Y}^-$&?&?&Y&Y\\
\hline
\multicolumn{5}{c}{}\\
\end{tabular}
\bigskip

Not surprisingly, the classes which cause the most difficulty are those in which $\tau$ has zeroes; we have a fairly complete understanding of whether or not solutions exist for the classes in which $\tau$ is bounded away from zero.

Near-CMC conformal data sets which are asymptotically Euclidean have also been studied. Techniques similar to those discussed briefly here show \cite{LCBY} that the criterion for the existence of solutions to the LCBY equations for near-CMC asymptotically Euclidean conformal data sets is very similar to that for CMC data.

\subsection{Far-CMC Data} 
\label{Far-CMC}
Without either the CMC or the near-CMC condition imposed, the analysis of the LCBY equations is considerably more difficult. For most sets of conformal data satisfying neither of these conditions (we use the label ``far-CMC" both for initial data sets $(\Sigma^3; \gamma_{ab}, K_{cd})$ and for conformal data sets $(\Sigma^3; \lambda_{ab}, \sigma_{cd}, \tau)$ which are not CMC and satisfy no smallness condition on $K^c_c$ or on $\tau$),  it is not known whether or not solutions exist.

There has, however, been recent work which begins to explore whether or not solutions exist for far-CMC data, and also explores their multiplicity.  We briefly  discuss three of these recent works here. 

The work of Holst, Nagy, and Tsogtgerel \cite{HNT09} together with the important follow-up work of Maxwell \cite{M05} effectively swaps the near-CMC condition of smallness of $|\nabla \tau |$ for a condition requiring the smallness of $|\sigma_{cd}|$. More specifically, their combined work shows that the LCBY equations admit a solution for a given set of conformal data on a closed manifold if the following conditions hold: (i) $\lambda_{ab} \in \mathcal{Y}^+$; (ii) $\lambda_{ab}$ does not admit a conformal Killing field; (iii) $|\sigma_{cd}|$ is sufficiently small; and (iv) $\sigma_{cd}$ is not identically zero. We emphasize the fact that these conditions include \emph{no} restriction on the mean curvature function $\tau$. We also note that while these conditions guarantee existence, they tell us nothing about uniqueness. 

Besides showing that a small but interesting class of far-CMC conformal data is mapped to solutions of the constraint equations, this work introduces a new analytical tool to the study of the conformal method. The proof that the LCBY  equations admit  solutions for sets of  conformal data of this sort described above relies on expressing solutions of the LCBY equation formally as fixed points of a map
\begin{equation}
\label{FixedPoint} 
\mathcal{F}: \phi \rightarrow \Delta^{-1}\big[ \frac{1}{8}R \phi -\frac{1}{8}(\sigma +L\{(\nabla\cdot L)^{-1} (\frac{2}{3}\phi^6 \nabla \tau)\})^2 \phi ^{-7} + \frac{1}{12} \tau^2\big],
\end{equation}
and proving that such fixed points exist using Schauder compactness. 

Schauder techniques also play a role in the work Dahl, Gicquaud and Humbert \cite{DGH}. They show that for certain classes of conformal data---data on a closed manifold $\Sigma^3$ with the metric (any Yamabe class) admitting no conformal Killing fields, with $\tau$ bounded away from zero, and with $|\sigma|^2$ not identically zero if $\lambda_{ab} \in \mathcal{Y}^-$--- solutions of the LCBY equations exist if the equation
\begin{equation} 
\label{limeq}
\Delta Y^a= |LY| \frac{1}{\tau}\nabla^a \tau
\end{equation}
does \emph{not} admit a solution. Thus, from this perspective, one proves the existence of LCBY solutions for a given set of conformal data by showing that \eqref{limeq} (called the \emph{limit equation} by the authors of \cite{DGH12}) admits no solution. This approach does not directly prove the existence of solutions of the LCBY equation for sets of far-CMC conformal data, but it can in principle  be used to do this. In \cite{GS12}, it shown that an analogous limit equation-type result holds for asymptotically hyperbolic conformal data sets. 

It is interesting that the derivation of these limit equation results relies on the study of solutions of a family of ``subcritical LCBY$_{\epsilon}$ equations" in which equation \eqref{Lichnerocoupled} is left unchanged, but \eqref{confmom} is replaced by 
\begin{equation}
\label{subcrit}
\nabla_m (LW)^m_a = \frac{2}{3}\phi^{(6-\epsilon)} \nabla_a\tau 
\end{equation}
It is much easier  to prove that solutions to the system \eqref{subcrit}-\eqref{Lichnerocoupled}, with $\epsilon>0$, exist, than to prove the existence of solutions to the LCBY equations. The idea then is to study the $\epsilon \rightarrow 0$ limit of solutions of the LCBY$_{\epsilon}$ system, and use the analysis of these limits to shed light on the LCBY system. Such an analysis leads to the limit equation results.

The third work concerning solutions of the LCBY equations for far-CMC conformal data which we discuss here is that of Maxwell \cite{M11}, in which he studies a very simple class of planar symmetric conformal data and finds somewhat surprising results. The conformal data sets he considers are characterized as follows: (i) $\Sigma^3 = T^3$; (ii) the metric is flat; (iii) $\sigma_{cd}$ has planar ($T^2$) symmetry, and therefore, as a consequence of the trace-free and divergence-free conditions, is a matrix with constant entries (two free constants); and (iv) $\tau$ is a step function, with a pair of discontinuities. Working with these simple data sets, which are  parametrized by a small set of constants, Maxwell can directly (numerically) construct solutions, if they exist. He finds that for certain ranges values of the parameters, no solutions exist; for other ranges, multiple solutions exist; and finally for a third range, unique solutions can be obtained. 

It is very intriguing to consider whether these results of Maxwell generalize to conformal data sets without any discontinuities,  and to data sets with less imposed symmetry. In any case, these results suggest that the behavior of the conformal method for far-CMC conformal data sets could be interesting and complicated. 

\section{The Conformal Thin Sandwich Method as an Alternative to the Conformal Method}
\label{CThin}

The conformal method has proven to be a remarkably useful tool for generating and parametrizing and analyzing solutions of the Einstein constraint equations.  It does, however, have some minor drawbacks:  
a) The conformal data is somewhat remote from the physical data, since the conformal factor changes the physical scale on different regions of space. b) While casting the constraints into a determined PDE  form has the advantage of producing PDEs of a relatively familiar (elliptic) form, one does give up certain flexibilities which are inherent in an underdetermined set of PDEs.  c) In choosing a set of conformal data, one has  to first project out  a divergence-free trace-free tensor field ($\sigma_{cd}$). d) While the LCBY system is conformally covariant in the sense discussed above in Section \ref{CMC-Closed} for CMC conformal data, this is not the case for non-CMC conformal data. 

The last two of these problems can be removed by modifying the conformal method in a way which York \cite{Y99} has called the ``conformal thin sandwich" (CTS) approach. The basic idea of the conformal thin sandwich approach is essentially the same as that of the conformal method. There are, however, two important differences. First, the CTS free data sets are larger than the free (conformal) data sets of the conformal method in the following sense: Like the conformal data sets, the CTS data sets include a conformal metric $\lambda_{ab}$ and a mean curvature scalar $\tau$. In addition, the CTS data sets include a trace-free tensor $U_{cd}$ to replace the divergence-free trace-free tensor $\sigma_{cd}$ of the conformal data sets, plus an extra scalar field $\eta$. Second, after solving the following  set of CTS equations (analogous to the LCBY equations) 
\begin{eqnarray}
\nabla_m((2\eta)^{-1} (LX))^m_a &=& \frac{2}{3}\Phi^6\nabla_a \tau + \nabla_m((2\eta)^{-1} U^m_a ),
\label{confmomCTS}
\\
\Delta\Phi &=& \frac{1}{ 8}R\Phi - \frac{1}{ 8}(U^{mn} +
LY^{mn})(U_{mn} + LY_{mn})\Phi^{-7}  
+ \frac{1}{ 12}\tau^2 \Phi^5,
\label{LichneroCTS}
\end{eqnarray}
for the conformal factor $\Phi$ and the vector field $Y^a$, one constructs not just the initial data $(\gamma_{ab}, K_{cd})$, but the lapse $N$ and the shift $M^a$ as well:
\begin{eqnarray} 
\gamma_{ab} &=& \Phi^4 \lambda_{ab} \\
K_{ab} &=& \Phi^{-2}(-U_{ab} + LY_{ab} ) +\frac{1}{3} \Phi^4 \lambda_{ab} \tau \\
N &=& \Phi^6 \eta \\
M^a &=& Y^a.
 \end{eqnarray}

It is clear that in using  the CTS approach, one need not project out a divergence-free part of a symmetric trace-free tensor. As well, one also readily checks that the CTS method is conformally covariant in the sense discussed above: the initial data $(\gamma_{ab}, K_{cd})$ and the lapse and shift $(N, M^a)$ generated from the CTS  data set  $( \lambda_{ab}, U_{ab}, \tau, \eta)$ and from the CTS data set $( \theta^4 \lambda_{ab}, \theta^{-2} U_{ab}, \tau, \theta^6 \eta )$ are identical. Furthermore, since the mathematical form of equations  (\ref{confmomCTS})-(\ref {LichneroCTS}) is very similar to that of (\ref{confmom})-(\ref{Lichnerocoupled}), the solvability results for the conformal method can be essentially carried over to the CTS approach. 
 
There is, however, one problematic feature of the conformal thin sandwich approach. The problem arises if we seek CMC initial data with the lapse function chosen so that the evolving data continues to have constant mean curvature\footnote{Such a gauge choice is often used in numerical relativity.}. In the case of the conformal method, after solving  (\ref{confmom})-(\ref{Lichnerocoupled}) to obtain initial data  $(\gamma_{ab}, K_{cd})$ which satisfies the constraints, one achieves this by proceeding to solve a linear homogeneous elliptic PDE for the lapse function. One easily verifies that solutions to this extra equation always exist. By contrast, in the CTS approach, the extra equation takes the form
\begin{equation}
\Delta(\Phi^7\eta) = \frac{1}{8} \Phi^7 \eta R +\frac{5}{2} (\Phi \eta)^{-1} (U-LY)^2  +\Phi^5 Y^m \nabla_m \tau - \Phi^5,
\end{equation}
which is coupled to the system (\ref{confmomCTS})-(\ref {LichneroCTS}). The coupling is fairly intricate; hence little is known about the existence of solutions to the system, and it has been seen that there are problems with uniqueness. These difficulties  do not arise, of course,  if one makes no attempt to preserve the constant mean curvature condition. 

\section{Gluing Solutions of the Constraint Equations}
\label{Glue}
Both the conformal method and  the conformal thin sandwich method are procedures for generating initial data sets which satisfy the Einstein constraint equations from scratch. The gluing procedures, which we discuss here, produce new solutions of the constraint equations by combining existing solutions.  While the gluing procedures have not yet turned out  to be as useful as the conformal method and CTS method  for the practical generation of physical interesting initial data sets, they have proven to be very effective for certain applications and for settling certain conjectures. 
We outline some of these applications below, after describing the two gluing procedures which have been developed, and how they work.

The \emph{asymptotic exterior gluing}, developed by Corvino and Schoen \cite{C00, CS03} works as follows. We presume that $(\Sigma^3, \gamma_{ab}, K_{cd})$ is an asymptotically Euclidean initial data set which satisfies the Einstein constraints, and also satisfies certain asymptotic conditions (as specified in \cite{CS03}). For any compact region
$\Xi^3 \subset \Sigma^3$ for which $\Sigma^3 \setminus \Xi^3 =
\mathbb{R}^3 \setminus B^3$ (where $B^3$ is a ball in $\mathbb{R}^3$),
there is a smooth asymptotically Euclidean solution of the constraints on $\Sigma^3$ which
is identical to the original solution on $\Xi^3 \subset \Sigma^3$, and
is identical to Cauchy data for the Kerr solution on $\Sigma^3 \setminus
{\tilde \Xi^3}$ for some ${\tilde \Xi^3} \subset \Sigma^3$. In words,
this technique allows one  to smoothly glue any interior region of an
asymptotically Euclidean solution to an exterior region of a slice of
a Kerr solution. There is generally a transition zone between the interior chosen region and the exterior Kerr region $\Sigma^3 \setminus \Xi$ which is unknown, but is a solution of the constraint equations. We note that for asymptotically Euclidean solutions of the
constraints with $K_{cd}=0$, this method glues any interior region to
an exterior region of a slice of the Schwarzschild solution.

In some sense, the Corvino-Schoen asymptotic exterior gluing result is very surprising. \emph{If} the constraint equations were a determined elliptic system, one would \emph{not} expect to be able to smoothly glue two solutions together like this, even with a transition region (satisfying the constraints). The key to proving that asymptotic exterior gluing indeed works is the exploitation of the underdetermined character of the constraints as a PDE system. Some of the ideas developed in the Corvino-Schoen work have also proven to be useful for localizing metric deformations as solutions of the constraints, as shown in the work of Chrusciel and Delay \cite{CD03}.

The other gluing procedure, \emph{connected sum gluing}, was developed first by Isenberg, Mazzeo and Pollack \cite{IMP01} with further work done together with Chrusciel \cite{CIP05} and with Maxwell \cite{IMaxP05}. The idea here is to start with a pair of solutions of the (vacuum) constraints $(\Sigma^3_1, \gamma_1, K_1)$ and $(\Sigma^3_2, \gamma_2, K_2)$ and to choose of a pair of points $p_1\in \Sigma^3_1$ and $p_2 \in \Sigma^3_2$, one point contained in each solution. Based on these solutions, connected-sum gluing 
 produces a new set of initial data $(\Sigma^3_{(1-2)}, \gamma_{(1-2)}, K_{(1-2)})$ with the following properties: i) $\Sigma_{(1-2)}$ is diffeomorphic  to the connected sum\footnote{The connected sum of these two manifolds is constructed as follows: First we remove a ball from each of the manifolds $\Sigma^3_1$ and $\Sigma^3_2$. We then use a cylindrical bridge $S^2\times I$ (where $I$ is an interval in $R^1$) to connect the resulting $S^2$ boundaries on each manifold.} $\Sigma^3_1\# \Sigma^3_2$; ii)  $(\Sigma^3_{(1-2)}, \gamma_{(1-2)}, K_{(1-2)})$ is a solution of the constraints everywhere on $\Sigma^3_{(1-2)}$; iii) On that portion of $\Sigma^3_{(1-2)}$ which corresponds to $\Sigma^3_1\setminus \{\textrm{ball around }  p_1\}$, the data $(\gamma_{(1-2)}, K_{(1-2)})$ is isomorphic to $(\gamma_1, K_1)$, with a corresponding property holding on that portion of $\Sigma^3_2$ which corresponds to $\Sigma^3_2\setminus \{\textrm{ball around }  p_2\}$.

Connected sum gluing can be carried out for fairly general sets of initial data. The sets may be asymptotically Euclidean, asymptotically hyperbolic, specified on a closed manifold, or indeed anything else. The only condition that the data sets must satisfy is that, in sufficiently small neighborhoods of each of the points at which the gluing is to be done, there do not exist nontrivial solutions $\xi$ to the equation
$D\Theta^{*}_{(\gamma, K)} \zeta=0$, where $D\Theta^{*}_{(\gamma, K)} $ is the operator obtained by taking the adjoint of the linearized constraint operator\footnote{If  a solution to this equation does exist on some region $\Lambda \in \Sigma^3$, it follows from the work of Moncrief  that the spacetime development of the data on $\Lambda$ admits a nontrivial isometry.}.  In work by Beig, Chrusciel and Schoen \cite{BCS05}, it is shown that this condition (sometimes referred to as ``No KIDs", meaning ``no (localized) Killing initial data") is indeed generically satisfied.

The details of the proof that connected sum gluing can be carried out as generally as described above are beyond the scope of this paper; see \cite{CIP05} along with the references cited in that work for a complete discussion. We do wish to note three features of the proof:  First, the proof is constructive in the sense that it outlines a systematic, step-by-step mathematical procedure for doing the gluing. In principle, one should be able to carry out the gluing procedure numerically. Second, connected sum gluing relies primarily on the conformal method, but it also requires the use of  a non-conformal deformation (dependent on the techniques of Corvino and Schoen, and of Chrusciel and Delay), so as to guarantee that the glued data is not just very close to the given data on regions away from the connecting bridge, but is indeed identical to it. Third, while the Corvino-Schoen asymptotic exterior gluing has not yet been proven to work for solutions of the constraints with source fields, connected sum gluing (up to the last step, which relies on Corvino-Schoen) has been shown to work for most matter source fields of interest \cite{IMaxP05}. It has also been shown to work for general dimensions greater than or equal to three. 

As noted above, while gluing has not seen wide-spread use as a procedure for producing physically interesting initial data sets, it has proven to be very valuable for a number of applications. We note a collection of these applications here:

\begin{enumerate}

\item {\it Spacetimes with Regular Asymptotic Structure:} Until recently, it was not known whether there is a large class of spacetime solutions  of the Einstein equations which admit the conformal compactification and consequent asymptotically simple structure at null and spacelike infinity characteristic of the Minkowski and Schwarzschild spacetimes. Using asymptotic exterior  gluing, together with Friedrich's analyses of spacetime asymptotic structures and arguments of Chrusciel and Delay \cite{CD02}, one can produce such a class of solutions.

\item{\it Initial Data for the Gravitational N-Body Problem:} To model the physics of a system consisting of N chosen astrophysical bodies interacting gravitationally, it is important to be able to construct initial data sets which solve the Einstein constraints and which accurately model the bodies of interest, their initial placement, and their initial momenta, all in a single asymptotically Euclidean space. Chrusciel, Corvino, and Isenberg \cite{CCI11} have used gluing techniques to show that for any chosen set of N asymptotically Euclidean solutions of the constraints representing black holes, stars, or other astrophysical objects of interest, one can construct a new asymptotically Euclidean solution which includes interior regions of these N chosen solutions, placed as desired (so long as the distances between the bodies are sufficiently large) and with the desired relative momenta.


\item {\it Adding a Black Hole to a Cosmological Spacetime:} Although
  there is no clear established definition for a black hole in a
  spatially compact solution of Einstein's equations, one can glue an
  asymptotically Euclidean solution of the constraints to a solution
  on a compact manifold, in such a way that there is an apparent
  horizon on the connecting bridge. Studying the nature of these solutions of the
  constraints, and their evolution, could be useful in trying to
  understand what one might mean by a black hole in a cosmological
  spacetime.
  
\item {\it Adding a Wormhole to Your Spacetime:}
  \index{wormhole!construction}  While we have
  discussed connected sum gluing as a procedure which builds solutions of the
  constraints with a bridge connecting two points on different
  manifolds, it can also be used to build a solution with a bridge
  connecting a pair of points on the {\it same} manifold. This allows
  one to do the following: If one has a globally hyperbolic spacetime
  solution of Einstein's equations, one can choose a Cauchy surface
  for that solution, choose a pair of points on that Cauchy surface,
  and glue the solution to itself via a bridge from one of these points
  to the other. If one now evolves this glued-together initial data
  into a spacetime, it will likely become singular very quickly
  because of the collapse of the bridge. Until the singularity
  develops, however, the solution is essentially as it was before the
  gluing, with the addition of an effective wormhole. Hence, this
  procedure can be used to glue a
  wormhole onto a generic spacetime solution. 
\item {\it Removing Topological Obstructions for Constraint
    Solutions:} We know that every closed three dimensional manifold
  $\Sigma^3$ admits a solution of the vacuum constraint equations. To show
  this, we use the fact that $\Sigma^3$ always admits a metric $\Gamma$ of
  constant negative scalar curvature. One easily verifies that the
  data $(\gamma=\Gamma, K=\Gamma)$ is a CMC solution. Combining this
  result with connected sum  gluing, one can show that for every closed $\Sigma^3$,
  the manifold $\Sigma^3 \setminus \{p\}$ admits both an asymptotically
  Euclidean and an asymptotically hyperbolic solution of the vacuum
  constraint equations.
  
\item{\it Proving the Existence of Vacuum Solutions on Closed
    Manifolds with No CMC Cauchy Surface:} \index{constant mean
    curvature!nonexistence}
  Based on the work of
  Bartnik  \cite{B88} one can show that if one has a set of
  initial data on the manifold $T^3\#T^3$ with the metric components
  even-reflective across a central sphere and the components of $K$ odd-reflective across
  that same central sphere, then the spacetime development of that
  data does not admit a CMC Cauchy surface. Using connected sum  gluing, one can
  show that indeed initial data sets of this sort exist \cite{CIP05}.

\end{enumerate}

\section{Comments on the Long-Time Evolution of Spacetime Solutions of Einstein's Equations}
\label{Evoln}

Once an initial data set satisfying the Einstein constraint equations has been obtained, one can evolve it into a spacetime satisfying the Einstein field equations. As guaranteed by the work discussed in Section \ref{Well-Posed}, there is a unique globally hyperbolic spacetime development of this initial data which contains (up to diffeomorphism) any other developments of the same set of data.

What do we know about the long-time properties of these maximal developments? The Hawking-Penrose singularity theorems \cite{HE73} tell us that (among spacetimes with a compact Cauchy surface), in one or the other direction in time, such developments ``generically" become causally geodesically incomplete, which means that there are causal geodesics in the spacetime which do not extend to infinite affine parameter length. This property of causal geodesic incompleteness is, however,  consistent with a wide variety of spacetime behavior, including curvature blowup, Cauchy horizon formation, and various topological anomalies \cite{ABBBGLSS}. 

One of the intriguing questions concerning spacetime developments is which of these behaviors---curvature blowup, Cauchy horizon formation, or something else---is expected to occur generically among those spacetimes which are causally geodesically incomplete. Penrose \cite{P99} has conjectured that curvature blowup is the generic behavior. This conjecture has been labeled the \emph{Strong Cosmic Censorship} conjecture (SCC)\footnote{The Strong Cosmic Censorships conjecture does not imply, and is not implied by, the Weak Cosmic Censorship conjecture, which concerns the generic formation of an event horizon around a singularity which forms in an asymptotically flat  solution of Einstein's equations.}  and it is viewed by many as one of the central questions concerning the evolutionary  behavior of solutions of Einstein's equations.

It is well-known that there are infinite dimensional families of solutions \cite{M84}  which have  bounded curvature and develop Cauchy horizons. The existence of these solutions  does \emph{not} disprove SCC. To formulate and study the SCC conjecture  carefully, one needs to define the notion of ``generic" in terms of the topology of the space of constraint-satisfying initial data sets on a fixed three dimensional manifold $\Sigma^3$, and then determine which sets of initial data evolve into a spacetime with unbounded curvature, and which do not. 

Strong Cosmic Censorship, formulated this way,  has been proven for certain families of solutions, most notably (by Ringstrom \cite{R05}) for the Gowdy family\footnote{The Gowdy family of solutions is introduced and characterized in \cite {G74} and  is studied extensively in \cite{C91}. SCC is proved for the Gowdy spacetimes in  \cite{R05}.}
which is characterized by the existence of a $T^2$ isometry group, and vanishing ``twists".  Numerical and other formal evidence strongly suggest that it is true for a wider class of spacetimes, with smaller isometry group \cite{IM02}. Proving or disproving this remains a major challenge. 

The Strong Cosmic Censorship conjecture concerns generic behavior \emph{among those spacetimes which are causally geodesically incomplete}. Distinct from this issue, and also of very significant interest is the question of which initial data sets evolve into spacetimes which extend an infinite (proper) time into the future and/or the past, and which do not. A complete answer to this question appears to be beyond our  current mathematical capabilities. However, as a small but very significant step towards answering this question, a number of researchers have focussed on the issue of the \emph{stability}---in terms of long time existence and structure---of solutions which exist for infinite proper time. 

The landmark stability result in general relativity is the proof by Christodoulou and Klainerman \cite{CK93} of the stability of Minkowski spacetime. They show, using energies based on the Bel-Robinson tensor to measure initial data perturbations, that the spacetime developments of initial data sets which are sufficiently close to Minkowski initial data do extend an infinite proper time into the future and into the past. Moreover, they show that these developments have the same asymptotic spacetime structure as Minkowski spacetime. These results have been extended to allow electromagnetic as well as gravitational initial data perturbations \cite{BZ09} and have also been strengthened in terms of the nature of the asymptotic structure which is shown to be stable \cite{BZ09}. 

If Minkowski spacetime is stable, one might logically proceed to consider if this is also the case for Schwarzschild spacetimes. However, since one knows that a small perturbation of Schwarzschild initial data which adds angular momentum will evolve into a Kerr solution rather than a Schwarzschild solution, one is led to consider the stability of Kerr spacetimes instead. The determination of whether or not Kerr spacetimes are stable is currently one of the most active areas of research in mathematical relativity. A recent report on the research directed  towards this goal appears in \cite{DR10}.

We note that the stability of other spacetimes has been established:  Friedrichs \cite{F86}  has shown that DeSitter spacetime is stable, Andersson and Moncrief \cite{AM04} have shown that Milne spacetime is stable, and Ringstrom \cite{R08} has shown that certain solutions of the Einstein-scalar field equations with accelerating expansion are stable. 

\section*{Acknowledgements}
This work was partially supported by NSF grant  PHY-0968612 at the University of Oregon.


\end{document}